\documentclass[preprint,aps,prl,superscriptaddress]{revtex4-1}

\usepackage{graphicx}
\usepackage{dcolumn}
\usepackage{bm}
\usepackage{amsmath}
\usepackage{latexsym}
\usepackage{hyperref}

\newcommand{\rfig}[1]{Fig.~\ref{#1}}

\makeindex

\begin{document}

\title{Electrical transport in nano-thick ZrTe$_5$ sheets: from three to two dimensions}

\author{Jingjing Niu}
\author{Jingyue Wang}
\affiliation{State Key Laboratory for Artificial Microstructure and Mesoscopic Physics, Peking University, Beijing 100871, China}
\author{Zhijie He}
\author{Chenglong Zhang}
\affiliation{International Center for Quantum Materials, Peking University, Beijing 100871, China}
\author{Xinqi Li}
\author{Tuocheng Cai}
\author{Xiumei Ma}
\affiliation{State Key Laboratory for Artificial Microstructure and Mesoscopic Physics, Peking University, Beijing 100871, China}
\author{Shuang Jia}
\email{gwljiashuang@pku.edu.cn}
\affiliation{International Center for Quantum Materials, Peking University, Beijing 100871, China}
\affiliation{Collaborative Innovation Center of Quantum Matter, Beijing 100871, China}
\author{Dapeng Yu}
\author{Xiaosong Wu}
\email{xswu@pku.edu.cn}
\affiliation{State Key Laboratory for Artificial Microstructure and Mesoscopic Physics, Peking University, Beijing 100871, China}
\affiliation{Collaborative Innovation Center of Quantum Matter, Beijing 100871, China}
\affiliation{Department of Physics, South University of Science and Technology of China, Shenzhen 518055, China}

\begin{abstract}
ZrTe$_5$ is a newly discovered topological material. Shortly after a single layer ZrTe$_5$ had been predicted to be a two-dimensional topological insulator, a handful of experiments have been carried out on bulk ZrTe$_5$ crystals, which however suggest that its bulk form may be a three-dimensional topological Dirac semimetal. We report the first transport study on ultra thin ZrTe$_5$ flakes down to 10 nm. A significant modulation of the characteristic resistivity maximum in the temperature dependence by thickness has been observed. Remarkably, the metallic behavior, occurring only below about 150 K in bulk, persists to over 320 K for flakes less than 20 nm thick. Furthermore, the resistivity maximum can be greatly tuned by ionic gating. Combined with the Hall resistance, we identify contributions from a semiconducting and a semimetallic bands. The enhancement of the metallic state in thin flakes are consequence of shifting of the energy bands. Our results suggest that the band structure sensitively depends on the film thickness, which may explain the divergent experimental observations on bulk materials.
\end{abstract}



\maketitle

\section{Introduction}
The topological insulator (TI) is a new quantum state of matter, which features a topologically protected metallic surface state with an insulating bulk\cite{Hasan2010,Qi2011}. Although the first experimentally demonstrated topological insulator, HgTe quantum well\cite{Konig2007}, is a two-dimensional (2D) one, the later found three-dimensional (3D) TIs, e.g., Bi$_2$Se$_3$ and Bi$_2$Te$_3$ families, have been studied the most. This is largely due to the easiness in materials growth and the large bulk band gap\cite{Chen2009,Xia2009}, which are advantageous in both research and technological point of view. However, the unintentional doping of the bulk, generating a sizeable parallel electrical conduction, hinders the understanding and application of the non-trivial surface states\cite{Analytis2010,Checkelsky2011,Xiu2011}. On the other hand, 2D TIs will not be affected by the problem owing to their gate tunability\cite{Pribiag2015,Qu2015}. Therefore, there have been great efforts in searching for new 2D TI materials\cite{Qian2014,Zhou2014,Si2014,Liu2015a} and some success has been made in heterostructures\cite{Pribiag2015,Qu2015,Du2015}. In light of the huge success of 2D crystals, it seems of great interest to find a 2D crystal TI material. Among many theoretical proposals, ZrTe$_5$ has attracted immediate attention after its prediction\cite{Weng2014,Li2016,Chen2015,Chen2015a,Zhou2015a,Yuan2015,Li2016a,Moreschini2016}, as it is believed to have a large band gap, $\sim 100$ meV, and it is a known material.

ZrTe$_5$ is a layered material in which layers are stacked in the crystallographic \textbf{b} direction. Each layer can be seen as ZrTe$_6$ prismatic chains in the \textbf{a} direction that are connected by zig-zag Te chains in the \textbf{c} direction. The material has been known for years, especially for its characteristic resistivity maximum at about 150 K\cite{Okada1980}. The research has just revived since the prediction of ZrTe$_5$ being a 2D TI\cite{Li2016,Chen2015,Chen2015a,Zhou2015a,Yuan2015}. However, the band structure remains unclear, as each among 3D Dirac semimetal\cite{Li2016,Chen2015,Chen2015a,Yuan2015}, weak TI\cite{Li2016a,Wu2016} and strong TI\cite{Manzoni2016}, have been favored by some experiments. So far, all experimental studies have been focused on bulk ZrTe$_5$ crystals, while no experiment on ultra thin ZrTe$_5$ sheets has been reported. Considering the prediction of a 2D TI for monolayer, it is important to see how the property evolves with decreasing thickness. Herein, we present experimental investigation on ultra thin ZrTe$_5$ sheets. A striking modulation of the resistivity maximum by thickness has been observed. In addition to the enhancement of metallicity with reducing thickness, the Hall resistance displays an evolution from nonlinear to linear, which clearly indicates shifting of energy bands. The ionic gating effect suggests presence of a semimetallic band, consistent with previous experiments on bulk. These observations coherently suggest that the band structure sensitively depends on the film thickness, consistent with theoretical calculations\cite{Weng2014,Fan2016}. The results provide an explanation for the divergent experimental observations on bulk materials.

\section{Experimental techniques}

ZrTe$_5$ single crystals were grown by an iodine vapour transfer method\cite{Okada1980}. The crystallographic structure was confirmed by X-ray diffraction and transmission electron microscopy (TEM). No trace of iodine impurities was observed in X-ray photoelectron spectroscopy. Raman spectra show the characteristics of ZrTe$_5$. Ultra thin sheets were prepared by mechanical exfoliation onto silicon substrates with 285 nm SiO$_2$. They can be readily identified by polarized light as the system is also quasi-one-dimensional due to the ZrTe$_6$ prismatic chain. The thickness of sheets was measured by
atomic force microscopy (AFM). Multi-electrode devices were prepared by e-beam lithography processes. Low temperature transport measurements were carried out using a standard lock-in method in a OXFORD variable temperature cryostat. Ionic gating experiment was performed using LiClO$_4$ (Alfa Aesar,0.3g) and PEO (Alfa Aesar,1g) mixed with anhydrous methanol (Alfa Aesar,15 mL) as the solid electrolyte\cite{Yu2015}.

\section{Results and discussions}

\subsection{Exfoliation and stability of thin ZrTe$_5$ flakes}
The interlayer bonding in ZrTe$_5$ is predicted to be as weak as graphite and much weaker than that in Bi$_2$Se$_3$ and Bi(111)\cite{Weng2014}. Mechanic exfoliation is supposed to work well. However, due to its quasi-one dimensional nature, sheets are easily torn apart along the $a$ direction. Consequently, it is harder to obtain large size flakes with uniform thickness by exfoliation compared with other isotropic 2D crystals. In our experiments, flakes were often narrow in the $c$ direction, making Hall measurements not always available. Moreover, the surface of flakes mostly exhibit steps, indicating thickness variation.

\begin{figure}[htbp]
\includegraphics[width=1\textwidth]{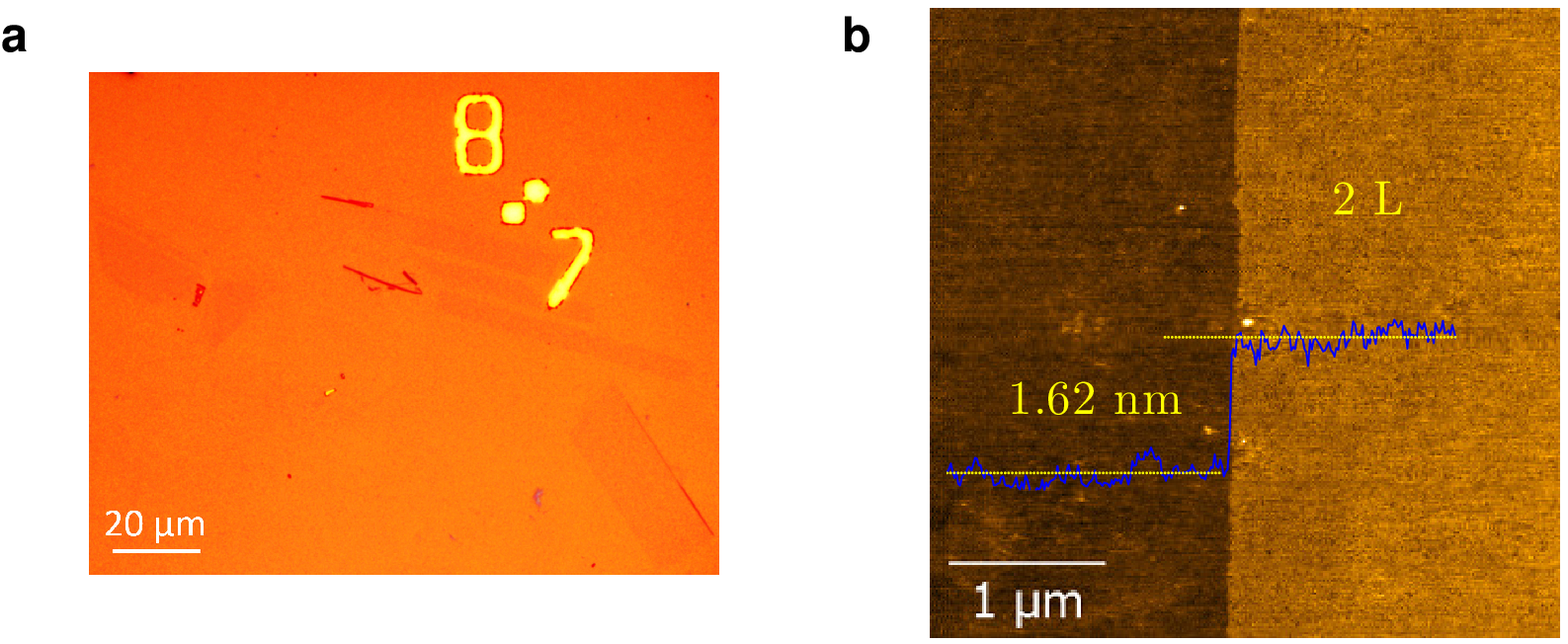}
\centering
\caption{\label{fig:s:bilayer} Large bilayer flakes by exfoliation. (a) Optical micrograph showing a few large size flakes. The contrast is enhanced to show the flakes. (b) AFM image of one of the flakes. The line profile (blue line) indicates a bilayer.}
\end{figure}

Nevertheless, in some cases, we could obtain single layer and bilayer flakes. \rfig{fig:s:bilayer}a shows the optical micrograph of a few large size bilayer flakes. One of them is 60 $\mu$m long. The largest one is about 20 $\mu$m by 40 $\mu$m, which really shows the potential to get large size few-layer flakes by exfoliation. The optical contrast for these thin flakes is extremely low so that it was hard to find them. An AFM image is shown in \rfig{fig:s:bilayer}b. From the line profile, we can read the thickness, 1.62 nm, suggesting a bilayer.

Usually, using polarized light, ZrTe$_5$ flakes can be readily identified under optical microscopy thanks to its quasi-one-dimension nature. This is done by shining a polarized light on a sample and detecting the reflected light of which the polarization is perpendicular to the incident light, \textit{i.e.}, crossed polarizers setup. When the $c$ axis of the sample is aligned at 45$^\circ$ with respect to the polarizers, strongest contrast can be obtained. Unfortunately, single layer and bilayer flakes didn't exhibit any contrast. No characteristic Raman peak was found for them. They were neither conducting. These observations led us to believe that they were oxidized.

\begin{figure}[htbp]
\includegraphics[width=0.9\textwidth]{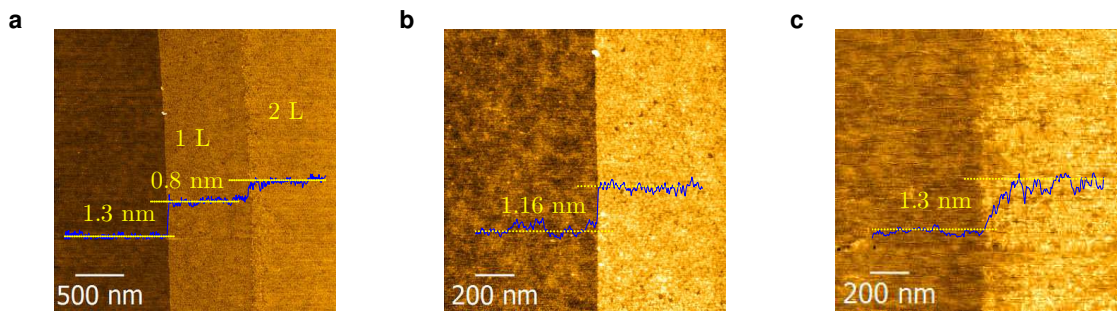}
\centering
\caption{\label{fig:s:oxidation} Time evolution of the surface morphology. (a) AFM image for a freshly cleaved flake. By the thickness, it can be seen that there are a single layer area and a bilayer area. (b) AFM image for another freshly cleaved single layer flake. The line profile indicates a similar surface roughness for the substrate and the flake. (c) AFM image for the sample in b after exposed in ambient air for 48 hours.}
\end{figure}

Oxidation is supported by the time evolution of the surface morphology. \rfig{fig:s:oxidation} shows the comparison of the surface morphology of flakes before and after being exposed in ambient air for 48 hours. Freshly cleaved flakes show relatively sharp edges. The rms surface roughness of the flakes is about 0.24 nm, similar to 0.2 nm for the substrate. After being exposed in air for 48 hours, the edge seemed substantially smeared and the roughness increased to 0.45 nm. Apparently, samples underwent significant changes.

\subsection{Structure characterization}
\begin{figure}[htbp]
\begin{center}
\includegraphics[width=1\columnwidth]{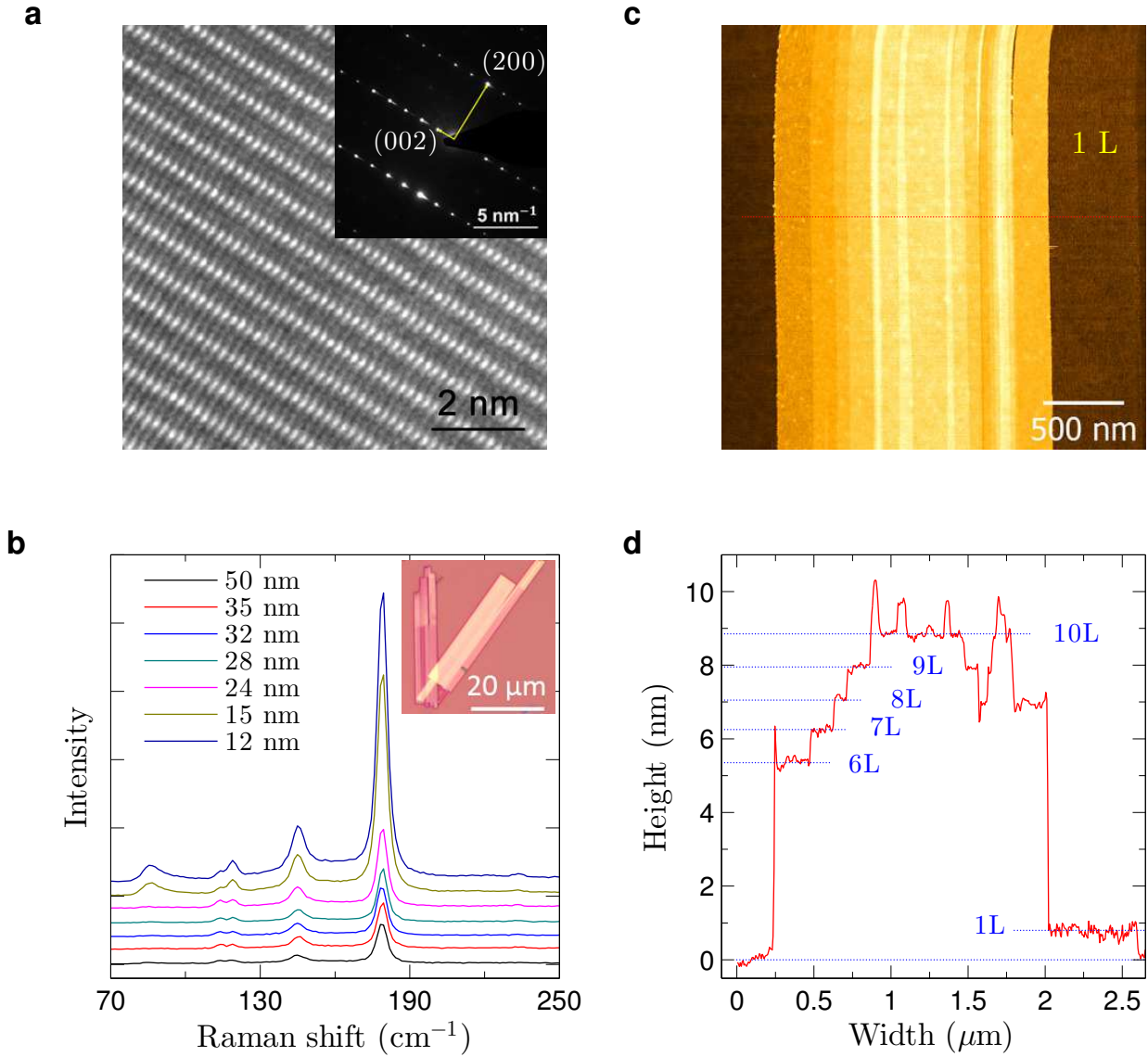}
\caption{Exfoliated thin ZrTe$_5$ flakes. (a) HRTEM image of a thin flake. Top-right, the electron diffraction looking down the [010] direction. The lattice constants are estimated as $a=0.400\pm0.002$
nm, $c=1.382\pm0.002$ nm. (b) Raman spectra at regions of different thickness, measured by AFM. Spectra are shifted for clarity. Inset, the optical image of the measured sample. (c) AFM image of a
flake. (d) The line profile shows steps, of which the height corresponds to the layer distance. On the right side, a single layer can be identified by the step height.}
\label{basic}
\end{center}
\end{figure}

\rfig{basic}a shows the high resolution TEM image of a thin sheet exfoliated from a ZrTe$_5$ crystal. The sharp diffraction spots indicate the high crystalline quality of the sample. Based on the diffraction pattern and the TEM image, we calculate the lattice constants of $a=0.400\pm0.002$ nm, $c=1.382\pm0.002$ nm. The interlayer distance $b/2$ is about $0.80\pm0.05$ nm estimated from the AFM results (\rfig{basic}d). Note that the interlayer distance determines the topological phase of ZrTe$_5$\cite{Fan2016,Manzoni2016}. Unfortunately, this cannot be done due to limited resolution of the AFM data. Raman spectra reproduce characteristic peaks reported for the material\cite{Taguchi1983a}. With decreasing thickness, the frequencies of the peaks remain unchanged, whereas the intensity is markedly enhanced. Such enhancement can be explained by an interference effect due to multi-reflection\cite{Wang2008}. However, the enhancement of the peak at 86 cm$^{-1}$ seems much stronger than others. This mode is connected with the vibrational mode of the Te zig-zag chain and it becomes stronger at low temperatures when the material displays a metallic behavior\cite{Landa1984}. This is very similar to our experiments, where enhancement was observed for thin flakes, which are more metallic. This will be shown later. We also want to point out that the peak at 115 cm$^{-1}$ is almost constant. The implications of these features are not clear and require further study. Instead, we concentrate on electrical transport. Since very thin flakes are not conducting due to oxidation, we limit our scope to sheets thicker than 10 nm.

\subsection{Transport}
\begin{figure}[htbp]
\begin{center}
\includegraphics[width=0.9\columnwidth]{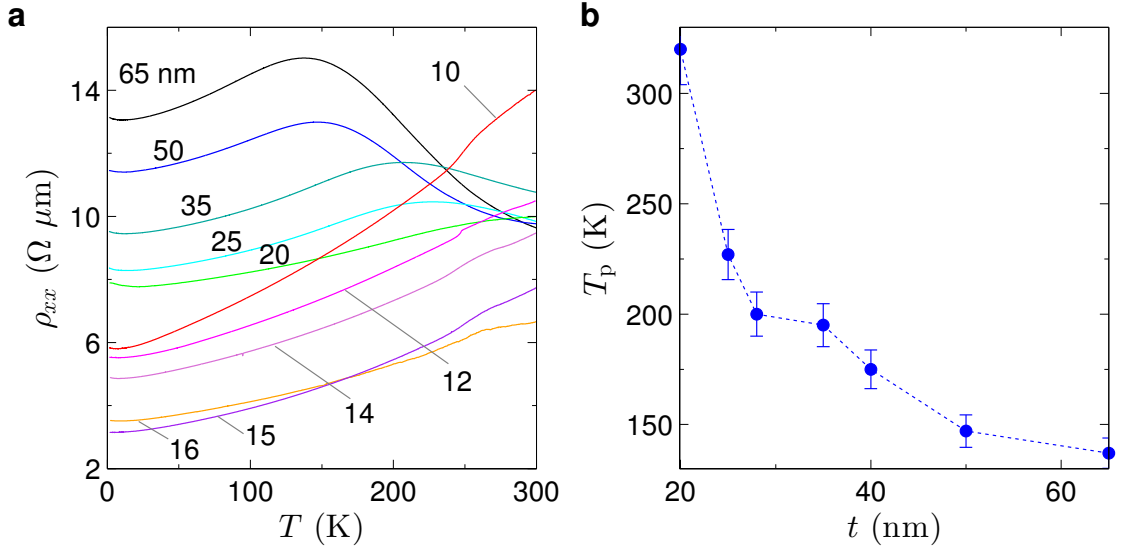}
\caption{Temperature dependence of resistivity for thin ZrTe$_5$ flakes. (a) Resistivity as a function of temperature for flakes of a series of thicknesses. For clarity, not all curves are shown. More
data can be found in the supplementary material. (b) Temperature of the maximum resistivity $T_\mathrm{p}$ versus $t$ for flakes thicker than 20 nm.}
\label{t.dep}
\end{center}
\end{figure}

The resistivity of the thick ZrTe$_5$ flakes as a function of temperature $T$ displays a maximum at about $T_\mathrm{p}=145$ K, consistent with the well-known resistivity anomaly in bulk crystals\cite{Okada1980}. However, as the thickness $t$ reduces, we observe a pronounced change of $T_\mathrm{p}$. It substantially shifts to a higher temperature, shown in \rfig{t.dep}a and b. The increase of $T_\mathrm{p}$ is remarkable, as it reaches 320 K at 20 nm. Below 20 nm, samples turn into metallic in the whole temperature range. The enhancement of metallicity is consonant with the change of the Raman mode at 86 cm$^{-1}$.

\begin{figure}[htbp]
\begin{center}
\includegraphics[width=0.85\columnwidth]{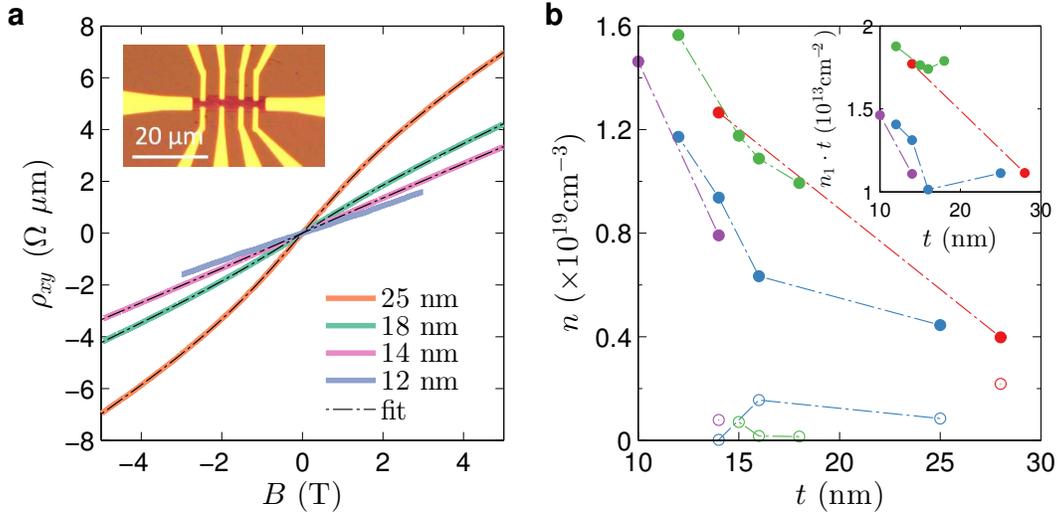}
\caption{Hall resistivity. (a) Hall resistivity as a function of magnetic field at 1.5 K for a few flakes of different thickness prepared in the same batch. The dashed lines are best fits to a two-band
model. (b) Carrier density as a function of $t$. Different colors represent different fabrication batches. Solid and open symbols represent $n_1$ and $n_2$, respectively. Since the Hall of the thinnest
sample in each batch is linear, the corresponding $n_1$ is directly calculated from the Hall slope.}
\label{hall}
\end{center}
\end{figure}

For some samples with larger size, we were able to fabricate Hall bars and measure the Hall resistance. A typical Hall bar structure is shown in the inset of \rfig{hall}a. For thicker samples, the Hall is nonlinear, seen in \rfig{hall}a, which strongly indicates carriers from more than one bands. This is not surprising as early studies on bulk have already found multiple bands\cite{Whangbo1982,Kamm1985}. Intriguingly, the nonlinearity gradually diminishes with reducing thickness. When the thickness is below 14 nm or so, it becomes linear. We have measured several batches of samples and the trend are well reproduced (see the supplementary information). Note that a linear Hall suggests dominance of a single band. Therefore, the evolution of the Hall implies a transition from multiple bands of carriers to single-band dominated carriers. Based on this observation, we have further carried out two-band fitting for the Hall. In a two-band model, the Hall resistivity
\begin{equation*}
\rho_{xy}(B)=\frac{B(B^2h_1h_2(h_1+h_2)+h_1{\rho_2}^2+h_2{\rho_1}^2)}{B^2(h_1+h_2)^2+(\rho_1+\rho_2)^2},
\label{hall.fit}
\end{equation*}
where $h_1$, $h_2$ are the hall coefficients and $\rho_1$, $\rho_2$ the resistivities for different bands, respectively. The total resistivity $\rho_{xx}$ satisfies $1/\rho_{xx}=1/\rho_1+1/\rho_2$, posing an additional constraint on the fitting parameters. From this fitting, the carrier density for each band, $n_1$ and $n_2$, are obtained by $n=1/(he)$, where $e$ is the elementary charge. The fitting results are shown in \rfig{hall} and Fig.~S3 in the supplementary material. In \rfig{hall}b, the carrier density is plotted as a function of $t$. To exclude the influence of the carrier density fluctuation among different bulk samples, we group the samples that were peeled off and patterned in the same batch. It is found that $n_1$ consistently increases with decreasing $t$, while $n_2$ remains small and becomes relatively negligible in thin samples.

One of the unique properties of topological materials is their non-trivial surface states. Some recent spectroscopy studies have shown indications of 2D Dirac surface states\cite{Li2016a,Wu2016,Manzoni2016}. So, it is tempting to relate $n_1$ to the surface. In that case, the 2D carrier concentration $n_1\cdot t$ should be more or less a constant. In the inset of \rfig{hall}b, $n_1\cdot t$ is plotted against $t$. A significant thickness dependence disfavors a surface origin. As we will explain later, $n_1$ is semimetallic and contributes to the metallic temperature dependence of resistivity below $T_\mathrm{p}$. If it is from the surface, its contribution in bulk materials would be too small to give rise to a substantial metallic behavior. Thus, we believe that $n_1$ is from the bulk. If we assume that the composition is independent of $t$, the change of $n_1$ can only result from a change in the band structure, \textit{i.e.}, shift of the band in energy. Shifting between two semiconducting bands is ruled out, because it leads to redistribution of carriers in two bands, which is inconsistent with no significant change in $n_2$. It is postulated that carriers $n_1$ reside in a semimetallic band resulting from band crossing and the crossing point shifts. The picture is consistent with a Dirac band observed in bulk materials\cite{Li2016,Chen2015,Chen2015a,Yuan2015}.

\begin{figure}[htbp]
\begin{center}
\includegraphics[width=1\columnwidth]{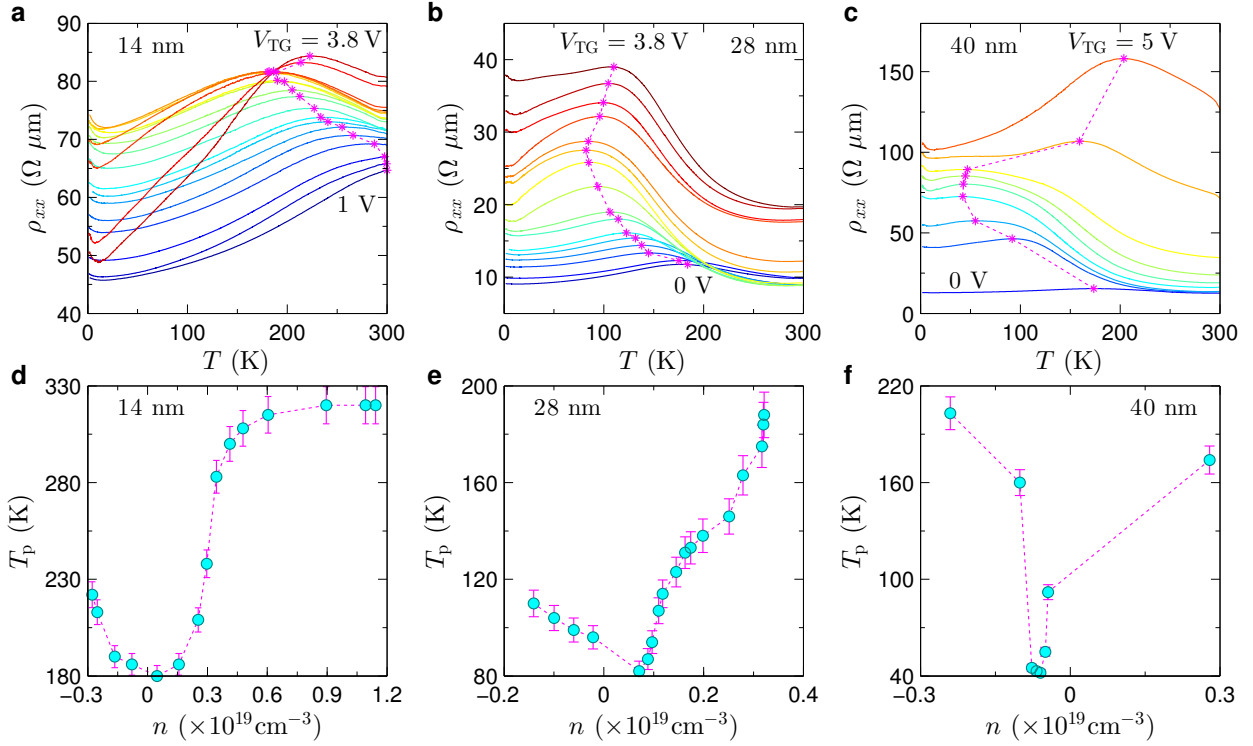}
\caption{Gate dependence of the resistivity peak. (a)(b)(c) Resistivity as a function of temperature at different gate voltages for three samples with $t=$14, 28 and 40 nm, respectively. In (b), the
four curves on the top are shifted up in $y$ axis so as to show the change of $T_\mathrm{p}$. In (c), the top two curves are shifted. The peak temperature $T_\mathrm{p}$ is marked by pink *. (d)(e)(f)
$T_\mathrm{p}$ versus $n$ for the samples in (a)(b)(c), respectively. $n$ is obtained from the low field Hall coefficient. In the vicinity of the transition, the Hall coefficient can be very small as it
must change its sign. Thus, $n$ cannot be correctly calculated. Instead, it is interpolated from the $n$ versus the gate voltage $V_\mathrm{TG}$ relatively away from the transition point (see
supplementary materials).}
\label{n.dep}
\end{center}
\end{figure}

Further evidence comes from gating experiments. The advantage of having ultra-thin samples is to be able to tune the carrier density by gating. We have performed back-gating and ionic liquid top-gating.
Similar results were obtained by both methods, except ionic gating offered a much larger carrier density range. Here, we mainly present the data obtained by ionic gating. The temperature dependent resistivity at different gate voltages for three samples with thickness of 14, 28 and 40 nm are plotted in \rfig{n.dep}a, b and c. With increasing gate voltage, $T_\mathrm{p}$ first decreases and then increases. The non-monotonic dependence has been observed in all samples. Hall measurements have also been carried out to obtain the carrier density. A sign reverse in Hall coefficient is observed, indicating a transition of carriers from holes to electrons. $T_\mathrm{p}$ is plotted against $n$ in \rfig{n.dep}d, e and f. For all three samples, the minimum $T_\mathrm{p}$ occurs close to the charge neutrality point. It is worthy to note that even at the charge neutrality point, the resistivity exhibits metallic temperature dependence at low temperatures, which strongly indicates semimetallicity.

\begin{figure}[htbp]
\begin{center}
\includegraphics[width=1\columnwidth]{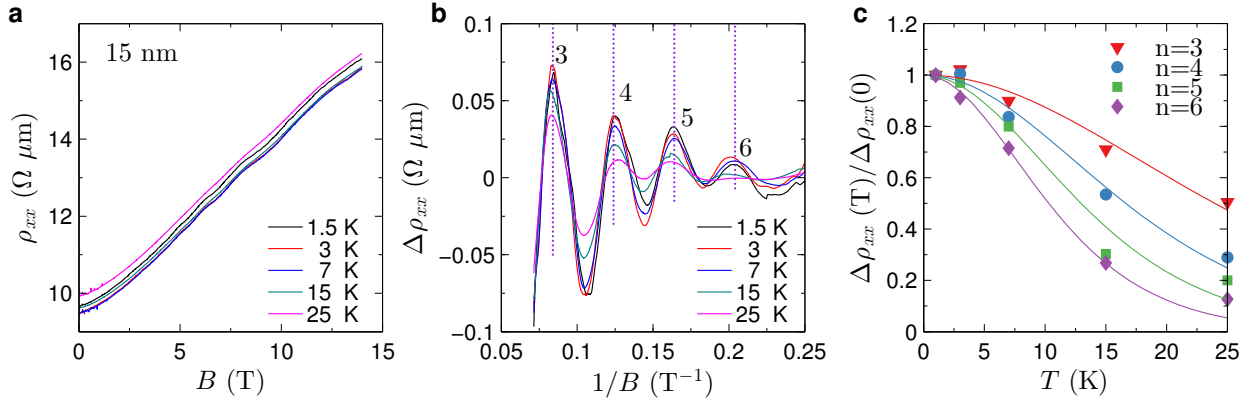}
\caption{Quantum oscillations of a 15 nm flake at $V_\mathrm{BG}=70$ V. (a) Magetoresistance. (b) Resistivity oscillations as a function of $1/B$ after subtracting a smooth background. (c) Damping of
the oscillation amplitude with temperature for Landau levels of $n=$3, 4, 5 and 6. Solid lines are fits to $R_\mathrm{T}$, see the text.}
\label{magneto}
\end{center}
\end{figure}

Magnetoresistance measurements reveal the nature of this semimetallic state. \rfig{magneto} shows the resistivity as a function of the magnetic field up to 14 T for a 15 nm thick sample. A backgate voltage of $V_\mathrm{BG}=70$ V was applied, so the carrier density is reduced. The sample shows a positivity magnetoresistance, which is often seen in Dirac materials. Similar positive magnetoresistance were observed in other samples, too. Interestingly, for this particular sample, small yet well-defined oscillations are discernible. After subtracting a smooth background, the oscillations are plotted against $1/B$, which display regular periodicity, indicating Shubulikov-de Haas oscillations. The damping of the oscillation amplitude $A$ with temperature is given by the Lifshitz-Kosevich equation,  $R_\mathrm{T}=A(T)/A(0)=\lambda/\sinh(\lambda)$, where $\lambda=2\pi^2k_\mathrm{B}Tm^*/\hbar eB$. Here, $k_\mathrm{B}$ is the Boltzmann constant, $e$ is the elementary charge and $m^*$ is the cyclotron mass. As depicted in \rfig{magneto}c, we have fit the temperature dependence of the amplitude to the equation of $R_\mathrm{T}$, yielding $m^*\sim 0.07 m_0$, where $m_0$ is the free electron mass. Assuming a linear dispersion, we have $\hbar k_\mathrm{F}=m^*v_\mathrm{F}$. The Fermi velocity is estimated as $5\times10^5$ m/s, in a good agreement with results reported by others\cite{Li2016,Chen2015,Yuan2015}. Our analysis of the quantum oscillations agrees well with experiments on bulk ZrTe$_5$\cite{Li2016,Chen2015,Chen2015a,Yuan2015}, indicating a massless Dirac band.

\subsection{Two-band model}
\begin{figure}[htbp]
\begin{center}
\includegraphics[width=0.8\columnwidth]{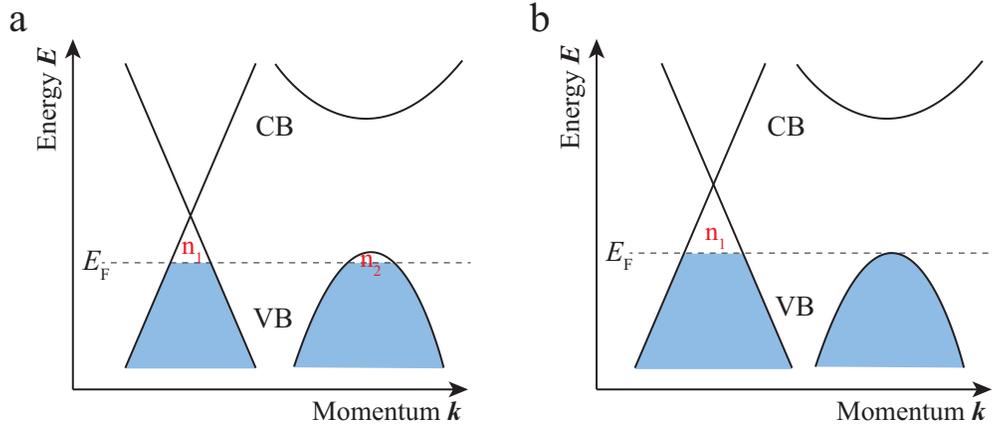}
\caption{Schematic two-band model. (a) For thicker flakes, two bands cross the Fermi level. One is a semimetallic, while the other is semiconducting. (b) For thin flakes, only the semimetallic band remains at the Fermi level due to band shifting.}
\label{two-band}
\end{center}
\end{figure}

Combining all the experimental observations, a consistent picture can now be formed, shown in \rfig{two-band}. From the thickness dependence of the Hall effect, it can be inferred that for thicker flakes, there are more than one bands at the Fermi level, which is in agreement with early studies on bulk crystals\cite{Whangbo1982,Kamm1985}. With decreasing thickness, the carrier concentration in one of the bands increases substantially. Consequently, the other band becomes negligible. The gate dependence of the Hall coefficient reveals a carrier type transition from hole to electron. Most importantly, there is no insulating state during the transition. Therefore, it can be concluded that one band is semimetallic. Analysis of the quantum oscillations is consistent with other groups' results that the carriers in the semimetallic band are indeed massless Dirac fermions.

The thickness dependence of the Hall effect indicates shifting of energy bands. It is interesting how the thickness affects the band structure. First, we note that recent work on other 2D crystals has discovered that the charge density wave transition can be significantly affected by thickness\cite{Goli2012,Yoshida2014,Xi2015}. Recent study on MoS$_2$ has shown that the interlayer distance increases with decreasing thickness, leading to reduction of interlayer coupling\cite{Cheng2012}. Second, the first principle calculation has found that the band structure is extremely sensitive to the lattice constants\cite{Weng2014,Fan2016}. In fact, experiments on bulk ZrTe$_5$ have found that the band shifts with temperature\cite{McIlroy2004,Li2016,Manzoni2015}. It is therefore not surprising to see such shifting caused by thickness. We speculate that the interlayer coupling is reduced in thin flakes, due to expansion in the layer distance. To test it, measurement of the lattice constants with precision is required in future study.

Such thickness dependent band structure may offer a hint on understanding the diverse experimental observations. It has been predicted that the band structure of ZrTe$_5$ sensitively depends on the interlayer spacing\cite{Weng2014,Fan2016}. With increasing lattice constants, it undergoes a topological transition from strong TI to an intermediate Dirac semimetal and then to a weak TI\cite{Fan2016}. If the actual lattice constant slightly varies with the growth method, temperature or thickness, the system will end up in different topological phases. The 3D Dirac semimetal is only a point in the phase diagram against the lattice constant, so its observation at first seems unlikely. However, the observed properties around this point can be close to those of the Dirac semimetal due to finite experimental resolution\cite{Manzoni2016}.

\begin{figure}[htbp]
\begin{center}
\includegraphics[width=0.6\columnwidth]{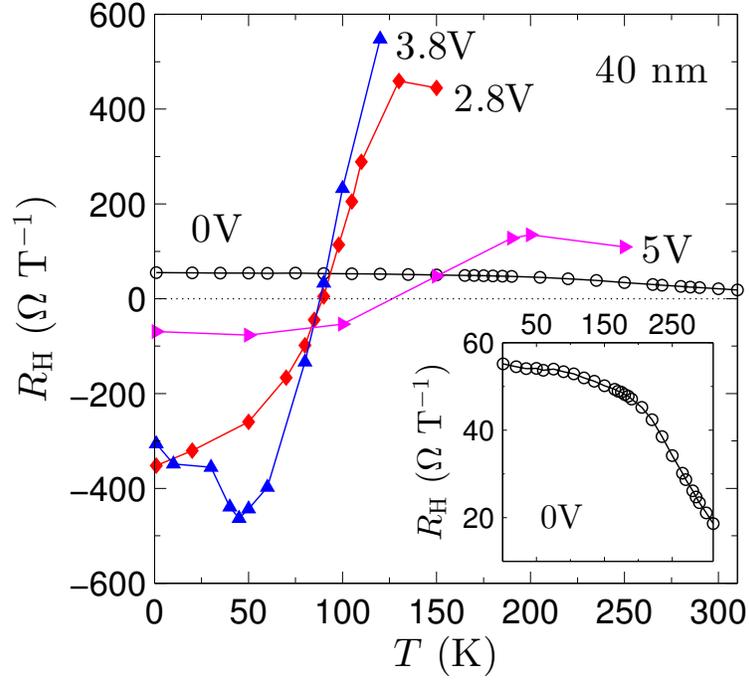}
\caption{Temperature dependence of low field $R_\mathrm{H}$ at different gate voltages for a 40 nm flake. Inset: re-plot of $R_\mathrm{H}$ at $V_\mathrm{TG}=0$ V.}
\label{Hall-T}
\end{center}
\end{figure}

At last, we discuss the origin of the resistivity peak, which has been a mystery for ZrTe$_5$. It was found that it coincided with a sign change of the Hall, which has led to proposals, such as charge density wave transition and polaronic conduction\cite{Izumi1982,Rubinstein1999b}. However, in our thin sheets that show the resistivity peak, the Hall remains positive up to room temperature, seen in \rfig{Hall-T}. This poses a strong constraint on possible models for the peak. We find that a two-band picture, one semimetallic band and one semiconducting band, naturally explains the peak feature. At low temperatures, the semimetallic band dominates the resistivity, giving rise to the metallic behaviour. With increasing temperature, the other semiconducting band takes over, due to either thermal activation or band shifting. The competition between two bands gives rise to a resistivity peak. Both bands can be of holes, as we observed in thin sheets. The change of the carrier density will alter the competition balance, hence the peak temperature, as demonstrated by the gating effect. Depending on the Fermi level, the system can shift from two hole bands to one hole band and one electron band, as shown in \rfig{Hall-T}. Then, the sign change of the Hall observed in bulk is restored.

\section{Conclusion}
In summary, we have studied the thickness and gate dependence of the transport properties of thin ZrTe$_5$ sheets. A strong modulation of the resistivity anomaly and a semimetallic behaviour have been observed. The Hall effect exhibits interesting dependence on the thickness. It is shown that these observations can be understood by a two-band model combined with band shifting upon thickness reduction. Our study offers a hint in understanding the divergent experimental observations on bulk.

\begin{acknowledgements}
We are grateful for enlightening discussion with X. Dai and N. L. Wang. This work was supported by National Key Basic Research Program of China (No. 2016YFA0300600, No. 2013CBA01603, No. 2012CB933404,
and No. 2016YFA0300802) and NSFC (Project No. 11574005, No. 11222436, and No. 11234001).
\end{acknowledgements}

\clearpage

\clearpage

\pagebreak
\widetext
\begin{center}
\textbf{\large Supplemental Materials: Electrical transport in nano-thick ZrTe$_5$ sheets: from three to two dimensions}
\end{center}
\setcounter{equation}{0}
\setcounter{figure}{0}
\setcounter{table}{0}
\setcounter{page}{1}
\makeatletter
\renewcommand{\theequation}{S\arabic{equation}}
\renewcommand{\thefigure}{S\arabic{figure}}
\renewcommand{\bibnumfmt}[1]{[S#1]}
\renewcommand{\citenumfont}[1]{S#1}

\begin{figure}[htbp]
\begin{center}
\includegraphics[width=0.45\textwidth]{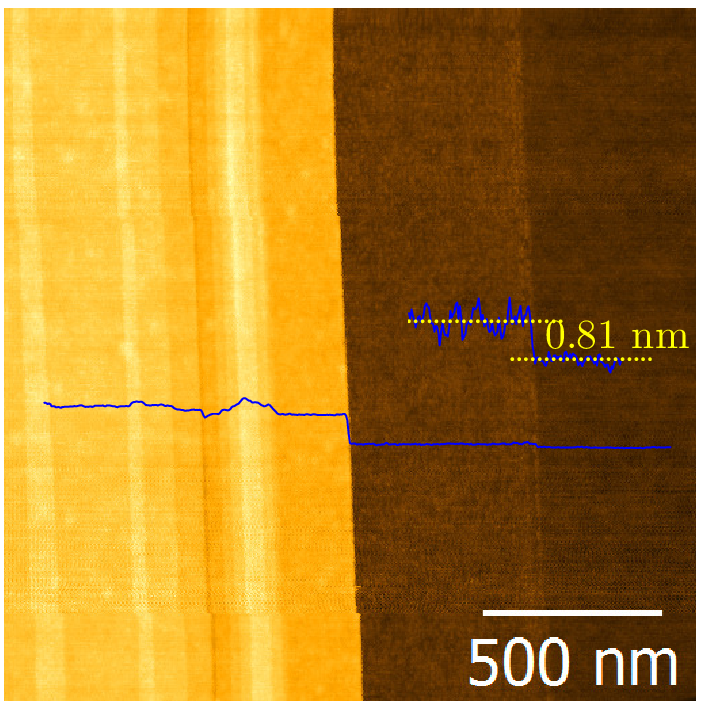}
\caption{\label{fig:s:afm1} Zoom-in AFM image for Fig.3d in the main text. The long blue line is a line profile. The short blue line is a zoom-in plot of the line profile, showing that the thickness of the stripe at the edge is 0.81 nm.}
\end{center}
\end{figure}

\newpage

\begin{figure}[htbp]
\includegraphics[width=0.45\textwidth]{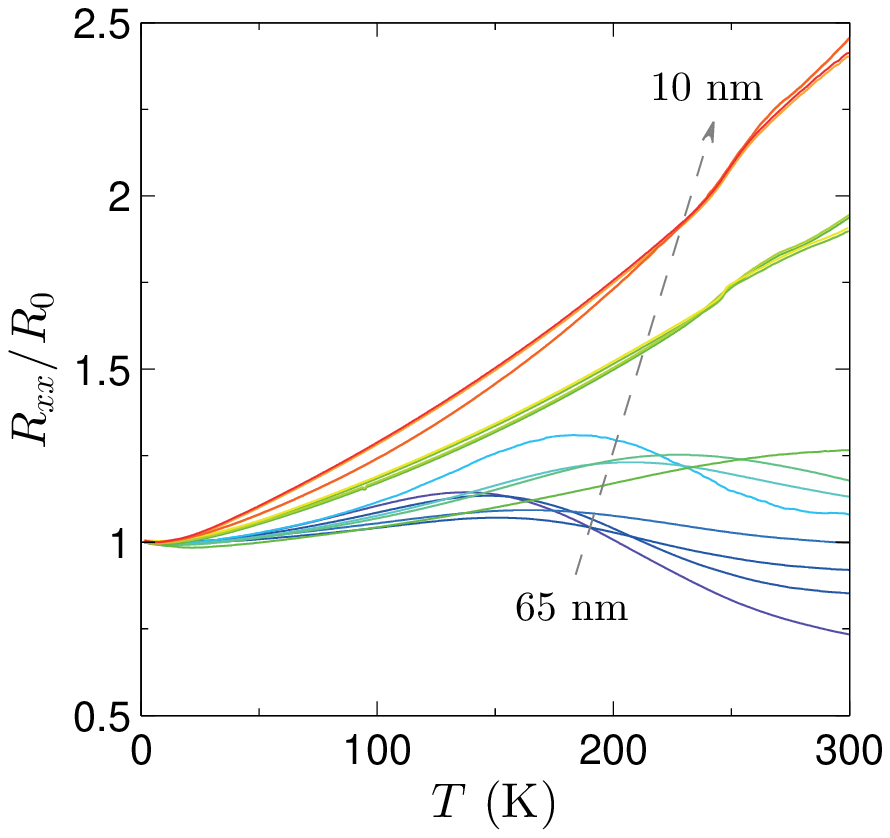}
\centering
\caption{\label{fig:s:t.dep} Temperature dependence of resistivity normalized to the low temperature resistivity $R_0$ for samples of thickness range from 65 to 10 nm.}
\end{figure}

The temperature dependence of resistivity for thicker samples ($t>20$ nm) shows clearly a hump at a certain temperature range from 145 K to 295 K, while for thin flakes (less than 15 nm) it exhibits a metallic behavior in the whole range of temperature below about 320 K. With decreasing thickness, a prominent shift of $T_\mathrm{p}$ from low to high temperature can be observed.

\newpage

\begin{figure}[htbp]
\includegraphics[width=0.9\textwidth]{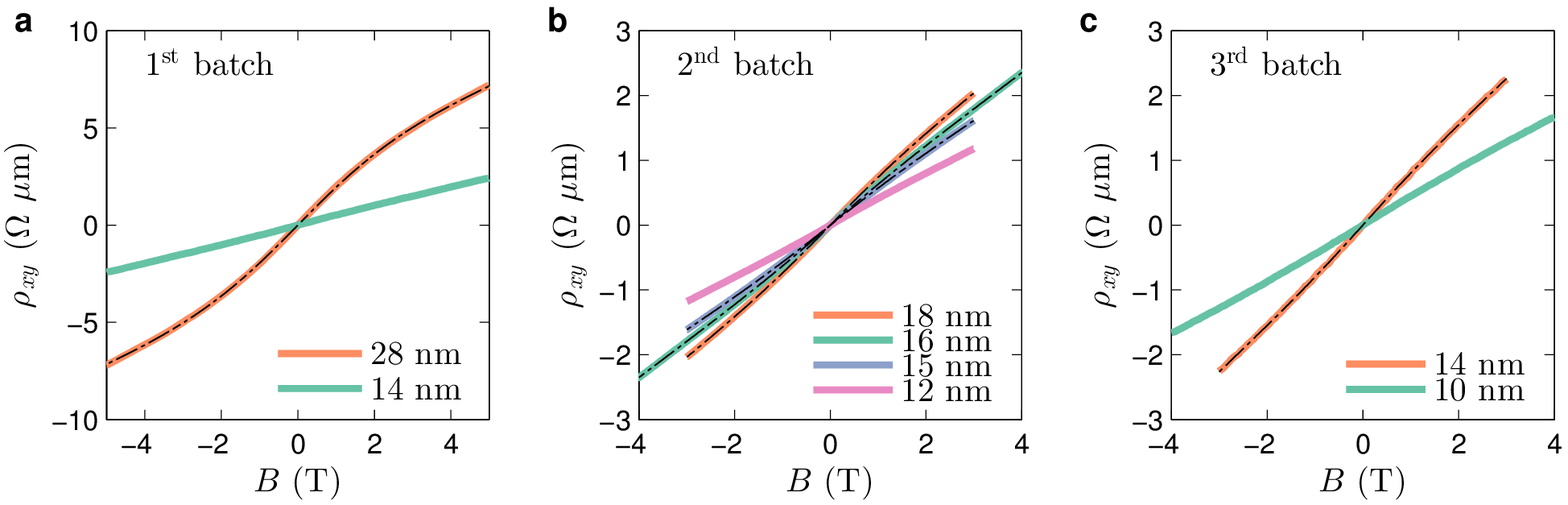}
\centering
\caption{\label{fig:s:hall} Hall resistivity $\rho_{xy}$ as a function of $B$ for flakes with different thickness. (a)(b)(c) are data for samples prepared in different batches. The dashed lines are best fits to a two-band model}
\end{figure}

The thickness dependence of the carrier density implies at least two bands involved for thick flakes. This is also supported by deviation of the Hall resistivity from linearity in higher magnetic fields. \rfig{fig:s:hall} shows the Hall resistivity $\rho_{xy}$ as a function of the magnetic field, they were well fitted with the two-band model. As explained in the text, we have grouped samples according to their batch numbers. When flakes are thick, $\rho_{xy}$ displays a nonlinear behavior. In contrast, when the thickness is less than about 15 nm, they are rather linear, indication of contribution from a single band. This trend appears in all batches, suggesting a transition from a two-band structure to a single band. This is consistent with the proposed band picture, \texttt{i.e.}, shifting one band away and leaving only one band at the Fermi level.

\newpage

\begin{figure}[htbp]
\includegraphics[width=0.9\textwidth]{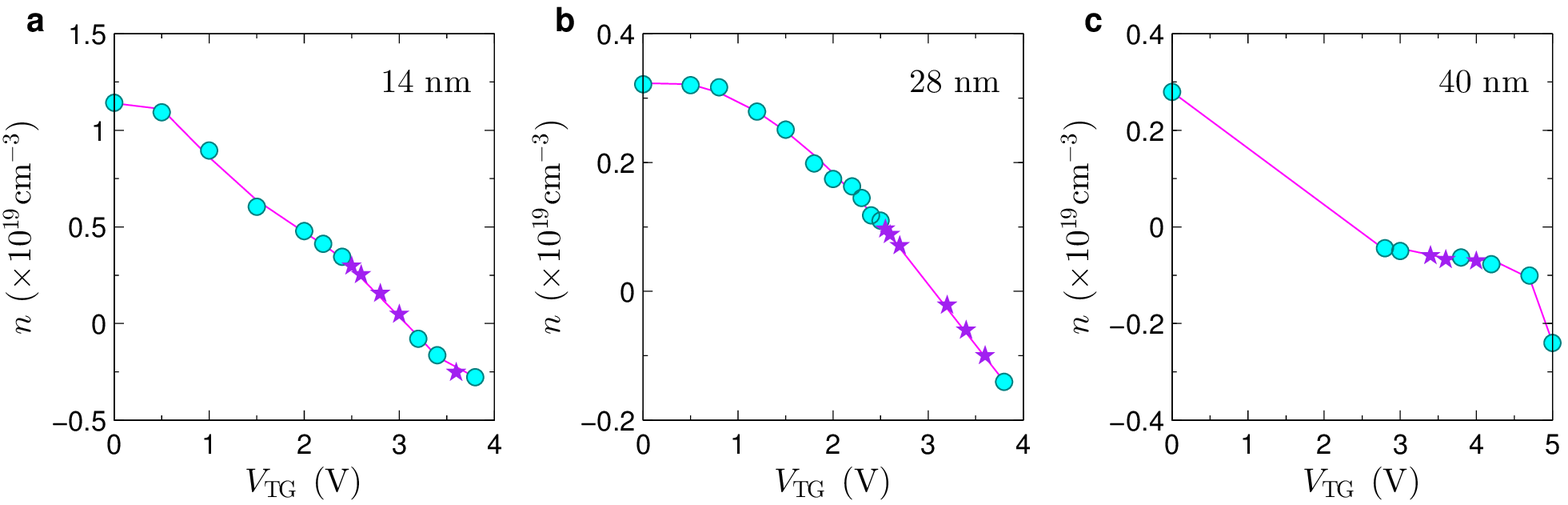}
\centering
\caption{\label{fig:s:density} Estimation of the carrier density $n$ at different gate voltages. (a) A 14 nm sample. (b) A 28 nm sample. (c) A 40 nm sample. Solid cyan circles are calculated from the Hall resistivity, while violet stars are linear interpolation.}
\end{figure}

The carrier density can be calculated from the Hall resistivity $\rho_{xy}$. The ionic gating experiments demonstrate a transition of carriers from holes to electrons, similar to graphene. Ideally, at the transition point, one should observe a divergent $\rho_{xy}$ and a sudden jump from positive to negative, as $\rho_{xy}$ is inversely proportional to the carrier density $n$. In reality, $\rho_{xy}$ increases to a finite value and then gradually reverses to negative due to density fluctuations. So, in the transition region, it is not appropriate to estimate $n$ from $\rho_{xy}$ any more. Therefore, we have obtained $n$ in this region by interpolation, as indicated in \rfig{fig:s:density}.

\newpage

\begin{figure}[htbp]
\includegraphics[width=0.9\textwidth]{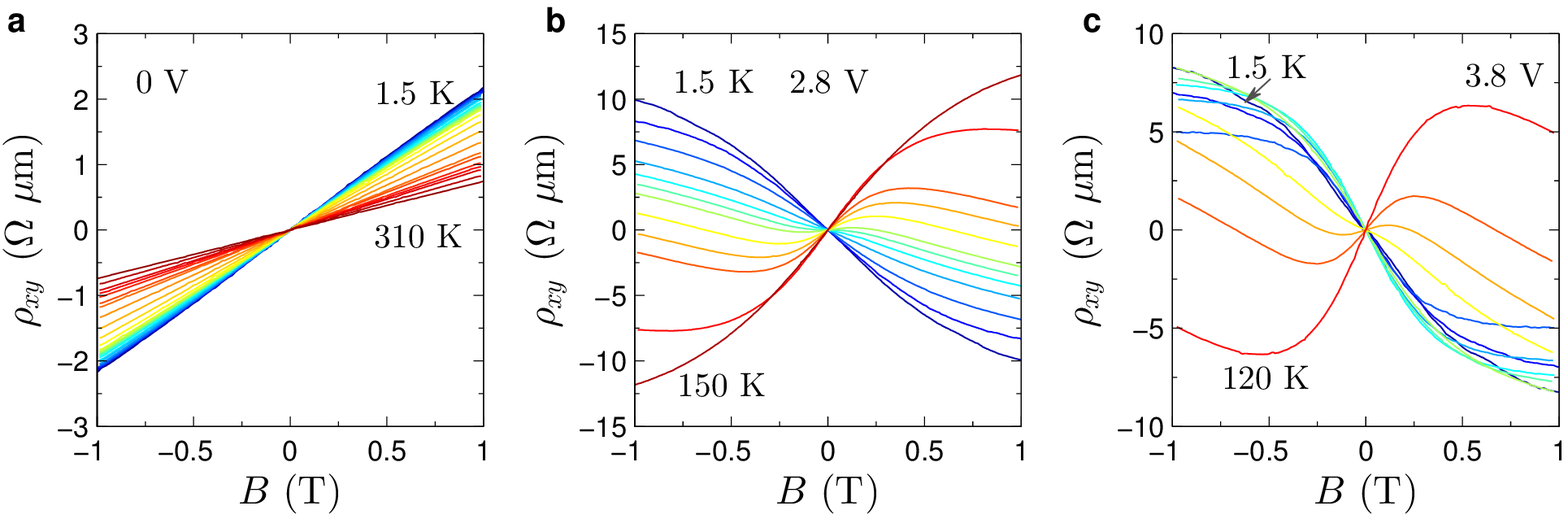}
\centering
\caption{\label{fig:s:Tdep} Evolution of the field dependence of the Hall resistivity with temperature for a sample of 40 nm thick at different gate voltages. (a) $V_\texttt{TG}=0$ V. $\rho_{xy}$ remains positive for the whole temperature range. (b) $V_\texttt{TG}=2.8$ V. $\rho_{xy}$ is negative at low temperatures and reverses its sign when temperature increases. (c) $V_\texttt{TG}=3.8$ V. Similar sign reversal is observed.}
\end{figure}

Early studies have found that the resistivity anomaly concurs with a sign change of the Hall resistivity and thermopower, indication of a change of the carrier type from electrons to holes. The observation led to speculations that the anomaly could result from the change of the band structure which also leads to the change of the carrier type[34,40]. However, our results show presence of the resistivity anomaly without change of the carrier type, see Fig.~9. We show that the competition between a Dirac semimetallic band and a semiconducting hole band can explain the resistance and Hall very well. \rfig{fig:s:Tdep} plots the Hall resistivity at $V_\texttt{TG}=$ 0, 2.8 and 3.8 V, from which the carrier density $n$ in Fig.~9 were obtained. When $V_\texttt{TG}=$ 2.8 and 3.8 V, it can be seen that $\rho_{xy}$ is negative at low temperature and become positive at high temperatures, similar to that of bulk material. Furthermore, $\rho_{xy}$ is strongly nonlinear, suggesting two bands.

\end{document}